\documentclass{aa}  

\usepackage{graphicx}
\usepackage[usenames,dvipsnames]{color}
\usepackage{pstricks}

\usepackage{txfonts}

\usepackage{hyperref}
\hypersetup{
    bookmarks=true,         
    unicode=false,          
    pdftoolbar=true,        
    pdfmenubar=true,        
    pdffitwindow=false,     
    pdftitle={P/2012 T1, X-SHOOTER},
    pdfauthor={C. Snodgrass et al.}, 
    pdfsubject={Planetary Science}, 
    pdfkeywords={},         
    pdfnewwindow=true,      
    colorlinks=true,        
    linkcolor=gray,         
    citecolor=blue,         
    filecolor=gray,         
    urlcolor=gray           
}

\begin{document}

   \title{X-shooter search for outgassing from Main Belt Comet P/2012~T1~(Pan-STARRS)\thanks{Based on observations collected at the European Organisation for Astronomical Research in the Southern Hemisphere
under ESO programme 290.C-5007(A)}}


   \author{C. Snodgrass
          \inst{1}
          \and
          B. Yang\inst{2}
          \and 
          A. Fitzsimmons\inst{3}
  }

   \institute{
  Planetary and Space Sciences, School of Physical Sciences, The Open University, Milton Keynes, MK7 6AA, UK         \\\email{Colin.Snodgrass@open.ac.uk}
         \and
             European Southern Observatory, Alonso de Cordova 3107, Vitacura, Santiago, Chile
         \and
            Astrophysics Research Centre, School of Mathematics and Physics, Queen's University Belfast, Belfast, BT7 1NN, UK
             }

   \date{Received ; accepted }

 
  \abstract
   {Main Belt Comets are a recently identified population of minor bodies with stable asteroid-like orbits but cometary appearances. Sublimation of water ice is the most likely mechanism for their recurrent activity (i.e. dust tails and dust comae), although there has been no direct detection of gas. These peculiar objects could hold the key to the origin of water on Earth.}
   {In this paper we present a search for the gas responsible for lifting dust from P/2012 T1 (Pan-STARRS), and review previous attempts at such measurements. To date such searches have mainly been indirect, looking for the common cometary gas CN rather than gasses related to water itself.}
   {We use the VLT and X-shooter to search for emission from OH in the UV, a direct dissociation product of water.}
   {We do not detect any emission lines, and place an upper limit on water production rate from P/2012 T1 of $8-9\times10^{25}$ molecules s$^{-1}$. This is similar to limits derived from observations using the {\it Herschel} space telescope.}
   {We conclude that the best current facilities are incapable of detecting water emission at the exceptionally low levels required to produce the observed activity in Main Belt Comets. 
   }


   \maketitle
%

\section{Introduction}

P/2012~T1 (Pan-STARRS) was discovered in October 2012 and immediately recognised as a potential new member of the recently identified class of Main Belt Comets \citep[MBCs,][]{Hsieh+Jewitt06}. It was the tenth candidate found and the third found by the Pan-STARRS survey \citep{Tholen12}. MBCs present a puzzle -- they have stable asteroid-like orbits, yet have comet-like appearances, implying the presence of volatile ices in their nuclei despite existing in a thermal environment that rules out surface ice. Although dust release can be caused by purely asteroidal processes -- e.g. collisions, rotational break-up, and/or binary merger \citep{Snodgrass10b,Stevenson12,Jewitt13d,Hainaut14} -- there is indirect evidence that some MBCs are genuine comets, with activity driven by sublimation of ice. The most convincing argument is the repeated activity seen in objects such as 133P/Elst-Pizarro and 238P/Read \citep{Hsieh10,Hsieh11b}, which is very difficult to explain by any other mechanism. Newer candidates, such as P/2012~T1, have not been known long enough to see if they repeat their activity each orbit, but morphological studies can be used to test whether a dust tail is due to prolonged activity or single events (like collisions). \citet{Moreno13} find that the morphology of P/2012~T1 is best explained by comet activity lasting a few months. However, there has not yet been a conclusive proof that MBCs are sublimation driven, which would be provided by a direct detection of gas in the coma of one of these comets.


In this paper we review previous attempts to detect outgassing from MBCs, before presenting our own VLT/X-shooter observations of P/2012 T1. We then discuss the implications of the non-detections, and compare this with expected activity levels.



\section{Searches for outgassing from MBCs}

Spectroscopy reveals the gaseous species in cometary comae via various strong fluorescence emission bands, mostly excited by solar UV radiation. In very bright comets it is possible to use high-resolution spectroscopy, and/or a wide range of wavelengths (ultraviolet [UV], visible light, infrared, sub-mm and radio), to get a very detailed measurement of the contents of the coma, even at an isotopic level \citep[e.g.][]{Bockelee-Morvan-cometsII}. From these, we infer the presence of volatile ices (known as the parent species) in the nucleus. As MBCs are very faint comets, there is little chance to detect any but the strongest emission lines. These are found in the UV/visible range; OH at 308 nm and CN at 389 nm are the easiest to detect in most comets. The OH band is the stronger of the two, but it is more difficult to observe from the ground due to absorption by ozone in the terrestrial atmosphere. Therefore CN is the species of choice for detecting outgassing from a faint comet, and this has been tried with a number of MBCs (Table \ref{tab:CNlimits}). None of these attempts were successful, despite using some of the world's largest telescopes, but allow upper limits on the gas production to be made. By assuming a `typical' CN:H$_2$O ratio for a comet, the limit on CN production is converted into a limit in the water production. The limits found all correspond to water production rates $Q({\rm H}_2{\rm O}) \le 10^{26}$  molecules per second (see Table \ref{tab:CNlimits}). This value would be quite low for a Jupiter family comet close to perihelion, but is less constraining at heliocentric distances $r \approx 3$ AU, where much lower activity levels are expected. Searches for CN have reached the sensitivity limit of current large telescopes, but do not rule out sublimation as the cause of the very low activity levels seen in MBCs -- these non-detections are not evidence for a lack of ice \citep{Jewitt12review}.

\begin{table*}
\caption{Previous upper limits on MBC gas production}
\begin{center}
\begin{tabular}{l l c c c c l}
\hline
MBC & Telescope & $r$ & Obs. date & $Q$(CN) & $Q$(H$_2$O) & Reference\\
 & & (AU) & & molec. s$^{-1}$ & molec. s$^{-1}$ & \\
 \hline
133P &  VLT &  2.64 & 2007/07/21 & $1.3 \times 10^{21}$ & $1.5 \times 10^{24}$ & \citet{Licandro11b} \\
176P* & Herschel & 2.58 &  2011/08/08 & -- & $4 \times 10^{25}$ & \citet{devalborro12}\\
324P & Keck & 2.66 & 2010/10/05 & $3 \times 10^{23}$ & $1 \times 10^{26}$ & \citet{Hsieh12c} \\
259P & Keck & 1.86 & 2008/09/30 & $1.4 \times 10^{23}$ & $5 \times 10^{25}$ & \citet{Jewitt09} \\
288P & Gemini & 2.52 & 2011/12/02 & $1.3 \times 10^{24}$ & $1 \times 10^{26}$ & \citet{Hsieh12b} \\
 & GTC & 2.52 & 2011/11/29 & $1.1 \times 10^{24}$ & -- & \citet{Licandro13} \\
596$^\dag$ & Keck & 3.1 & 2010/12/17 & $9 \times 10^{23}$ & $1 \times 10^{27}$ & \citet{Hsieh12a} \\
P/2013 R3 & Keck & 2.23 & 2013/10/01 & $1.2 \times 10^{23}$ & $ 4.3 \times 10^{25}$ & \citet{Jewitt14a}\\
313P & Keck & 2.41 & 2014/10/22 & $1.8 \times 10^{23}$ & $6 \times 10^{25}$ & \citet{Jewitt2015g}\\
P/2012 T1 & Keck & 2.42 & 2012/10/19 & $1.5 \times 10^{23}$ & $5 \times 10^{25}$ & \citet{Hsieh13} \\
  & Herschel & 2.50 & 2013/01/16 & -- & $7.6 \times 10^{25}$ & \citet{ORourke13} \\
 \hline
\end{tabular}
\end{center}
* 176P was not visibly active (no dust release) at the time of the Herschel observations.\\
\dag{} The dust ejected from (596) Scheila was almost certainly due to a collision, rather than cometary activity \citep[e.g.][]{Bodewits11,Ishiguro11a,Yang11}.
\label{tab:CNlimits}
\end{table*}%


Moreover, theoretical work suggests that only water ice can survive in an MBC, and more volatile ices (such as HCN, which is thought to be the parent species for CN) would have been lost. Thermal models which calculate the heat conduction into the interior of asteroids find that water ice can survive in MBC-like orbits when buried at depths of around 100 m, but that the entire interior of a km-scale body will reach a temperature above the HCN sublimation point \citep{Prialnik09,Capria12}. While it is possible to imagine that more volatile species survive as gas trapped within water ice, they certainly cannot be retained in large amounts, and will definitely not have the same relative abundance as other comets.

This means that the assumption of a similar CN:H$_2$O ratio for MBCs is probably incorrect, and searches for CN gas, as an indicator of ice sublimation, may never succeed, no matter how sensitive. Instead the search must focus on direct detection of water or its dissociation products (e.g. OH). Two attempts have been made using the {\it Herschel} space telescope to look for water directly (176P -- \citet{devalborro12}; P/2012 T1 -- \citet{ORourke13}), but both provided only upper limits (Table \ref{tab:CNlimits}). In the first case the observers were also unlucky that 176P did not return to activity when predicted -- no dust was seen in visible wavelength images taken at around the same time as the {\it Herschel} observations \citep{Hsieh14}. For P/2012 T1 there was clearly visible dust activity at the time of the {\it Herschel} observations, but no gas was seen. This result is discussed further in section \ref{sec:discussion}. 


\section{X-shooter observations of P/2012 T1}

\subsection{Observations}

Our approach, rather than trying to detect water directly, is to search for the strong UV emission band from OH(0-0) at 308 nm. The photodissociation of H$_2$O $\rightarrow$ OH+H in cometary comae is well understood, and can therefore be used to obtain reliable water production rates from OH line strengths with a few assumptions. The OH band is also the strongest in typical comet spectra \citep[e.g.][]{Feldman04}, and should therefore produce a clear signal in even low activity comets. The difficulty with this band is the strong absorption in the terrestrial atmosphere due to ozone at UV wavelengths, which is around 80\% at 308 nm. Good skies, large telescopes and very sensitive instrumentation are therefore required to recover the remaining photons, and many modern low-to-medium resolution spectrographs do not even try to get to such blue wavelengths (VLT/FORS, for example, cuts off below 350 nm by design).

X-shooter is designed to provide medium-to-high resolution spectroscopy over a very wide wavelength range in one shot, using three arms optimised for UVB, visible and NIR wavelengths \citep{X-SHOOTER}. Its bluest order covers 300--320 nm, making it potentially sensitive to OH emission lines in comets. Combined with the collecting power of the 8 m VLT, detection of low production rate cometary outgassing should be possible with X-shooter. The very wide wavelength range (up to 2.5 $\mu$m) also gives another advantage of X-shooter for studying comets -- many other potential emission lines can also be investigated simultaneously, including CN (even if is not expected in MBCs), and the continuum from the dust can give clues about its composition. The NIR arm covers the range where solid state absorption from ice grains in the coma could potentially be detected, at 1.5 and 2.0 $\mu$m.
For example, \citet{Snodgrass2016} used X-shooter to observe 67P/Chuyumov-Gerasimenko around the time of the {\it Philae} landing, although the low activity level of the comet at the time (at $r$ = 3 AU) and poor visibility (airmass $\sim$ 2) meant that only upper limits on gas production rates could be measured.

\begin{table}
\caption{Observations details}
\begin{center}
\begin{tabular}{l c c c c c}
\hline
UT Date & $r$ & $\Delta$ & $\alpha$ & airm. & $N_{\rm obs}^\dag$\\
 & (AU) & (AU) & (deg) & & \\
\hline
2012/12/14 & 2.46 & 1.67 & 17 & 1.2 & 4\\
2012/12/18 & 2.47 & 1.71 & 18 & 1.2 - 1.4 & 6\\
\hline
\end{tabular}
\end{center}
\dag{} Number of observations. Each consists of parallel exposures of 900s duration in UVB and NIR arms, and 855s in VIS.
\label{tab:obs}
\end{table}%

Following the discovery of P/2012 T1 and initial visible wavelength photometry to assess its total brightness and dust production, we applied for and were awarded four hours of Director's discretionary time to try these challenging X-shooter observations. The observations took place in service mode during dark time in December 2012, within three months of the initial discovery. Details on the observing geometry are given in Table \ref{tab:obs}. We used slit widths 1.0, 0.9 and 0.9 arcsec in the UVB, VIS and NIR arms, respectively. We also observed the solar analogue star SA93-101 immediately after the comet on both nights.

\subsection{Data reduction}

\begin{figure}
\includegraphics[width=\columnwidth,trim=40 35 85 50,clip]{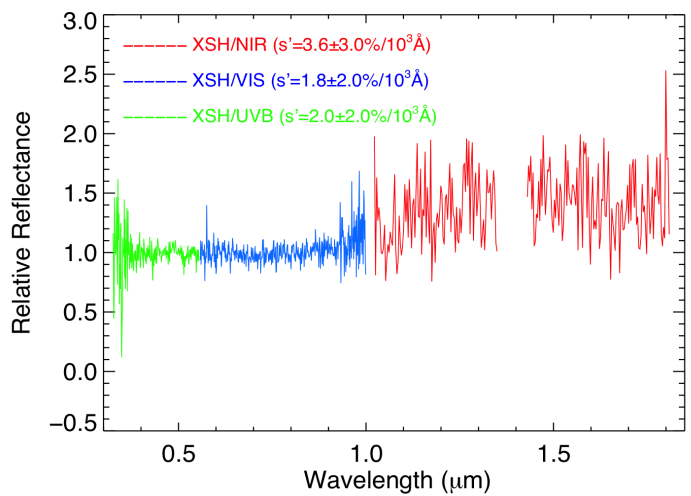}
\caption{Combined X-shooter reflectance spectrum of P/2012 T1. The UVB and VIS spectra have been binned into intervals of 10\AA\ and the NIR spectrum has been binned into intervals of 60\AA\ to increase the signal-to-noise. The two spectral regions, centred at 1.4 and 1.9 $\mu$m, that are heavily contaminated by the telluric absorptions have been cut out.}
\label{fig:spec-full}
\end{figure}

We first processed the data using the X-shooter pipeline \citep{XSHOOTER-pipeline}, which performed the following steps: bias-subtraction, cosmic ray detection and removal, flat-fielding, wavelength-calibration and order-merging. Our target was rather faint, especially in the bluest orders of the UVB arm, where the signal level was very low. In order to get reliable signal, we extracted the 1D spectrum from a two-dimensional image using the {\tt apall} procedure from IRAF. We only applied flux calibration to the UVB section, where the important OH and CN emission lines are located. We removed telluric absorption features as well as the solar gradient via dividing the comet spectrum by that of the solar analogue star, SA93-101. The resulting reflectance spectrum is shown in fig.~\ref{fig:spec-full}.

\subsection{Dust continuum}

The reflectance spectrum has a remarkably flat slope, in agreement with the Keck spectrum published by \citet{Hsieh13}, and in contrast to the red slopes typically seen in most comets \citep[e.g.][]{Hadamcik09}. The neutral slope of the dust spectrum matches the nucleus spectrum measured for other MBCs \citep{Licandro11b}, assuming that the coma and nucleus spectra can be compared. Observations at 67P showed that the total coma spectrum matched the nucleus \citep{Snodgrass2016}, but the dust size distribution and grain surface composition can influence coma reflectance properties \citep[e.g.][]{Kolokolova2004}. The comet spectrum appears rather weak in the NIR, especially so in the K-band. We therefore present the NIR portion of the continuum in the J and H band only. Although the dust slope appears slightly redder at  wavelengths $> 1 \mu$m (4 $\pm$ 3 \%/1000\AA), it is consistent, within the uncertainty, with the neutral slopes found at shorter wavelengths (2 $\pm$ 2 \%/1000\AA), with no evidence for absorption features due to ice or minerals. Again there is a contrast with typical comets, where the spectral slope changes between the visible and NIR \citep[e.g.][]{Snodgrass2016}.


\begin{figure}
\includegraphics[width=\columnwidth,trim=40 35 85 50,clip]{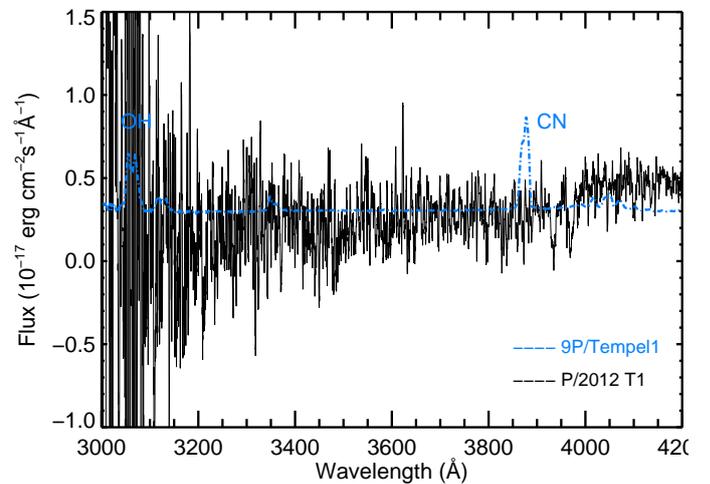}
\caption{Combined flux spectrum of P/2012 T1 (without any solar spectrum removal), zoomed in on the UV region around the expected OH and CN emission bands, which has been smoothed over 20 adjacent points (about 3.7\AA). A rescaled and offset spectrum of 9P/Tempel 1, obtained by \cite{Meech11}, is overlaid to illustrate the locations of the OH(0-0) and CN(0-0) emission bands.}
\label{fig:spec-zoom}
\end{figure}

\subsection{Search for gas emission}
%
%
%

Figure~\ref{fig:spec-zoom} shows the combined, flux calibrated, spectrum (without removing the dust continuum) in the region where strong cometary emission bands are expected. We do not detect any emission above the noise level.
The fluorescence scattering efficiency (g-factor) of OH is a strong function of heliocentric velocity due to the complex solar spectrum in this region. Using the heliocentric radial velocity of P/2010 T1 of $+1.9$ km sec$^{-1}$ at this time and the corresponding g-factor calculated by \citet{Schleicher+AHearn88} of $g[{\rm OH(0-0)}]=1.74\times10^{-15} r^{-2}$ ergs s$^{-1}$ molecule$^{-1}$, our upper limit on the flux corresponds to a total number of OH molecules within the X-shooter slit of $N({\rm OH})<4.8\times10^{27}$.

To derive the corresponding upper limit to the sublimation rate requires the use of a coma density model. To facilitate comparison with other comet studies, we first use the common Haser model \citep{Haser57}. We adopt the parent and daughter scalelengths used by \cite{Ahearn95} and a nominal parent velocity given by $v_p=0.85r^{-0.5}=0.6$ km s$^{-1}$ \citep{Cochran+schleicher93}. Numerically integrating the resulting theoretical column densities within our slit implies a total production rate of OH of $Q({\rm OH})<6\times10^{25}$ molecules s$^{-1}$. As our observations occurred near solar maximum we use the active Sun branching ratio of 0.80 \citep{Huebner92}, giving a production rate of $Q({\rm H_2O})<8\times10^{25}$ molecules s$^{-1}$ from the Haser model.

Although this result is directly comparable with many other analyses, the analytical Haser model contains several unphysical assumptions. Therefore we have also calculated  an upper limit to $Q({\rm H_2O})$ using a version of the Monte Carlo model first described by \cite{Combi+Delsemme80}. We use a Maxwellian velocity distribution for the parent H$_2$O molecules, and we assume ejection only from the sunward hemisphere with a distribution proportional to $\cos \theta$, where $\theta$ is the subsolar latitude. The true photodissociation lifetimes of H$_2$O and OH are dependent on the variable FUV and solar H-$\alpha$ flux. From solar indices measured during our observations and the relationships given by \cite{Budzien94}, we use lifetimes against dissociation at 1 AU of $\tau({\rm H_2O})=8.9\times10^4$ s and $\tau({\rm OH})=1.5\times10^5$ s. An analysis of previous ground-based measurements of outflow velocities was performed by \cite{Tseng07}, but these were all for comets  with  $Q({\rm H_2O}) \geq 10^{28}$ molecules s$^{-1}$.  Instead we utilise the measurements made by the MIRO instrument onboard Rosetta when it arrived at 67P in August 2014 and the nucleus was in a low-activity state. \cite{Lee15} found a clear correlation between the terminal expansion velocity and production rate of H$_2$O, with $v_p\simeq0.73$ km sec$^{-1}$ for production rates of $Q({\rm H_2O})\geq 10^{25}$ molecules s$^{-1}$ when the comet was at $r\simeq3.5$ AU. Assuming a $v_p\propto r^{-0.5}$ relationship as expected from sublimation theory, we therefore assume $v_p=0.8$ km s$^{-1}$ for P/2012 T1 at the time of our observations. With these parameters and again using a branching ratio of 0.8, we obtain $Q({\rm H_2O})<9\times10^{25}$ molecules~s$^{-1}$.

We have performed similar modelling for our upper limit to the CN $\Delta \nu=0$ flux. With a g-factor of $3.12\times10^{-13} r^{-2}$ ergs s$^{-1}$ molecule$^{-1}$ \citep{Schleicher10}, the number of CN molecules in our slit is $N({\rm CN})<4.0\times10^{24}$. Haser modelling then gives a production rate of $Q({\rm HCN})<3\times10^{22}$ molecules s$^{-1}$, assuming all CN is a photodissociation product of HCN.  For the Monte Carlo model we use the same outflow velocity as before of 0.8 km s$^{-1}$ and photodissociation lifetimes from \cite{Huebner92} for an active sun of $\tau({\rm HCN})=3.2\times10^4 r^2$~s and $\tau({\rm CN})=1.35\times10^5 r^2$~s. Using these parameters resulted in the same upper limit of $Q({\rm HCN})<3\times10^{22}$ molecules s$^{-1}$.


\section{Discussion}\label{sec:discussion}


\subsection{Comparison with previous searches for gas emission}

There are two previously published attempts to detect gas emission from P/2012 T1: a spectrum covering the CN region using the Keck telescope \citep{Hsieh13}, and a search for water using the ESA {\it Herschel} space telescope \citep{ORourke13}. Neither attempt was successful. \citet{Hsieh13} derived an upper limit of 
$Q({\rm CN})<1.5\times10^{23}$ molecules s$^{-1}$ using a Haser model. Assuming a ratio  $Q({\rm CN})/Q({\rm OH)}=3\times10^{-3}$ as found in normal comets \citep{Ahearn95} and a branching ratio of 0.9 this corresponded to $Q({\rm H_2O})<5\times10^{25}$ molecules s$^{-1}$.  Our improved limit on CN production from our X-shooter observations, assuming the same composition, implies a water production limit of $Q({\rm H_2O})<1\times10^{25}$ molecules s$^{-1}$. This  improvement is largely due to the significantly longer total integration time used at the VLT compared to the Keck observations, and UVB arm of X-shooter being very efficient around the CN band. 

The {\it Herschel} observation has the unique advantage of searching directly for water, and found a limit of $Q$(H$_2$O) $< 7.63 \times 10^{25}$ molecules s$^{-1}$ \citep{ORourke13}. Although higher than the result from CN, there are less assumptions involved in this upper limit  and we therefore regard it as the best previous limit on water production from a MBC.  The limit on water production derived from our X-shooter search for OH emissions is $Q({\rm H_2O})<9\times10^{25}$ molecules s$^{-1}$, similar to the {\it Herschel} value. Additionally, the {\it Herschel} observations took place approximately one month after our VLT observations, due to pointing constraints to keep the telescope behind its sun-shade, where according to the lightcurve of \cite{Hsieh13} activity had decreased since the epoch of our observations.


\begin{figure}
\includegraphics[width=\columnwidth,trim=40 35 85 80,clip]{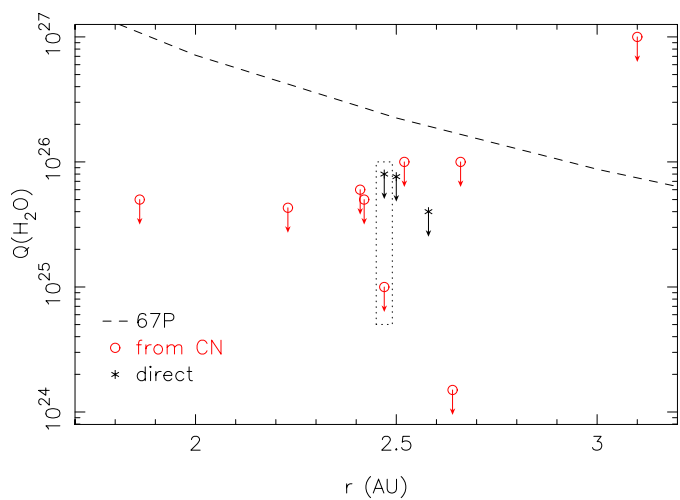}
\caption{Upper limits on water production, via CN or direct (H$_2$O or OH). Water production rate for 67P, as a typical JFC, shown by dashed line for comparison (empirical fit to inbound data -- \citealt{Hansen2017}). Our X-shooter results are highlighted by the dotted box.}
\label{fig:Qplot}
\end{figure}

In Fig.~\ref{fig:Qplot} we show these various upper limits, along with the other attempts for MBCs listed in table~\ref{tab:CNlimits}. It is clear that all upper limits fall around the same production rate, with a weak trend towards more distant objects having higher (less constraining) limits due to their relative faintness and the g-factor's $g \propto r^{-2}$ dependence. The most constraining limits come from our X-shooter CN measurement, which had a longer integration time, and the limit for 133P published by \citet{Licandro11b}, which was derived from a column density through the outer coma region and a vectorial model, and therefore may not be directly comparable. This plot largely shows the limit of sensitivity of the current generation of large telescopes. 

Figure~\ref{fig:Qplot} also shows the water production rate for 67P as a `typical' JFC for comparison, which was higher at main belt distances, demonstrating that these upper limits at least show MBCs to be very weakly active comets.  However, it is worth noting that attempts to detect CN in 67P from the ground at large $r$ inbound also produced only upper limits, with a similar sensitivity to the MBC searches \citep{Snodgrass2016}: The comparison between ground-based observations and {\it Rosetta} measurements  shows that the CN/H$_2$O ratio varies significantly, even within one apparition for a single comet. 


\subsection{Comparison with dust activity}
As there are no reported detections of gas associated with MBCs to date, we use the observed dust coma to estimate the possible sublimation rate from P/2012 T1. \cite{Hsieh13} report an absolute R-band magnitude of $m_R(1,1,0)\simeq 15.5$ within a 5 arcsec radius aperture near the epoch of our observations, equivalent to a comet dust activity index of $A f \rho\simeq 10$ cm. \citet{Moreno13} measured a fainter $m_R(1,1,0)\simeq 17.3$  with a smaller aperture, however as they do not report the aperture size we do not use their data. Assuming a typical cometary dust particle albedo of $A=0.04$ gives a total dust cross section within the aperture of  $6\times10^7$ cm$^2$. Taking a mean particle diameter of $\sim10^{-6}$ m and assuming a dust outflow velocity slower than the gas of $\sim 0.1$ km s$^{-1}$ implies a dust production rate of  $\sim 7\times10^{19}$ particles s$^{-1}$.  With an assumed density of 1000 kg m$^{-3}$ this gives a mass loss rate in dust at this time of $\sim0.6$ kg s$^{-1}$.

Using sophisticated modelling of the dust coma and tail, \citet{Moreno13} calculate a dust mass loss rate at the time of our observations $\sim 90$ days after perihelion of $\sim 0.2-0.3$ kg s$^{-1}$, within a factor 2--3 of the simplistic calculation above. As P/2012 T1 was observed near perihelion, we assume that H$_2$O sublimation was the main driver of activity. The estimated gas/dust mass loss ratio varies significantly between normal comets. If we assume a nominal ratio of $\sim 1$ for P/2012 T1, this gives a water sublimation rate of $Q({\rm H_2O})\sim 1-2\times10^{25}$ molecules s$^{-1}$. This is  below both our upper limit of $Q({\rm H_2O})$ and that of \cite{ORourke13}, but by less than a factor of 10. This strongly indicates that direct detection of OH at future perihelion passages may be possible if P/2012 T1 exhibits the same level of activity. Importantly, we also note that this rules out a very high gas/dust ratio of $\geq 10$ for this MBC. It is possible that P/2012 T1 has a gas/dust ratio $<$ 1, as was observed by Rosetta at 67P \citep{Rotundi2015Sci}, which would imply a true gas production further below our derived limit.


Assuming the normal $Q({\rm CN})/Q({\rm OH)}$ and branching ratios, our upper limit for water production based on our CN upper limit would be $Q({\rm H_2O})\sim 1 \times10^{25}$ molecules s$^{-1}$. This is of the same order as the dust mass-loss rate. As we do not detect CN, we are led to conclude that this object has a similar or lower $Q({\rm CN})/Q({\rm H_2O})$ than normal comets, or that $Q({\rm gas})/Q({\rm dust}) < 1$, or some combination of both. While this potentially supports the models of \citet{Prialnik09} as discussed above, it is clear that significantly deeper integrations are required to further investigate this. 

From our upper limit to $Q({\rm OH})$, we can use standard sublimation theory to estimate an upper limit to the effective actively sublimating area. We assume a C-type taxonomy with a geometric albedo of $0.06$ \citep{Mainzer11} and a phase integral of $0.3$, giving  a Bond albedo for the surface of $\sim 0.018$. Taking the vapour pressure and latent heats for H$_2$O from \cite{Prialnik04},  bare  ice would  sublimate at a rate of  $1.8\times10^{21}$ molecules s$^{-1}$ m$^{-2}$. Therefore our non-detection of OH, the upper limit of $Q({\rm H_2O})<9\times10^{25}$ molecules s$^{-1}$ corresponds to a bare ice patch of area $\leq 5\times 10^4$ m$^2$. This in turn corresponds to 0.2\% of surface area if the nucleus radius is 1.3 km, which is the upper limit from photometry taken when the comet was active \citep{ORourke13}. The active patch could be as much as 4\% of the surface if the true size of the nucleus is as small as smallest known MBC, 259P/Garradd (radius = 300 m; \citealt{MacLennan+Hsieh2012}). These active fraction areas are comparable with those found for typical comets \citep{Ahearn95}.



\section{Conclusions}

Unfortunately, all of the available upper limits are consistent with the predicted production rates required to drive the observed dust production, and are therefore not constraining on theory -- water ice sublimation remains the most likely explanation for the repeated activity of MBCs, but at a level below below current detection limits. X-shooter limits on water production via OH are similar to previous results using CN as a proxy, but have the advantage of not relying on an unknown CN/H$_2$O ratio for MBCs. Our upper limits may be within one order of magnitude of detecting outgassing, based on some reasonable assumptions, but we will require a brighter MBC to make the first direct gas detection with current technology. In a few years, the launch of the {\it James Webb} space telescope will give a powerful direct way to study water outgassing from comets; it is expected to be sensitive enough to detect the water escaping from MBCs \citep{kelley16}. 

\begin{acknowledgements}

We thank the ESO DG and DDT panel for awarding time to this project, and the service mode observer, telescope operator and USD staff who enabled us to get the data.
We thank Dan Bramich for useful advice on the X-shooter pipeline. 
CS received funding from the European Union Seventh Framework Programme (FP7/2007-2013) under grant agreement no. 268421 and from the UK STFC in the form of an Ernest Rutherford Fellowship.
AF acknowledges support from UK STFC grant ST/L000709/1. 
This paper was written, in part, at the International Space Science Institute (ISSI) in Bern, Switzerland. We thank ISSI for their support.

\end{acknowledgements}

\bibliographystyle{aa}
\bibliography{mbc}

\begin{thebibliography}{51}
\expandafter\ifx\csname natexlab\endcsname\relax\def\natexlab#1{#1}\fi

\bibitem[{{A'Hearn} {et~al.}(1995){A'Hearn}, {Millis}, {Schleicher}, {Osip}, \&
  {Birch}}]{Ahearn95}
{A'Hearn}, M.~F., {Millis}, R.~L., {Schleicher}, D.~G., {Osip}, D.~J., \&
  {Birch}, P.~V. 1995, Icarus, 118, 223

\bibitem[{{Bockel{\'e}e-Morvan} {et~al.}(2004){Bockel{\'e}e-Morvan},
  {Crovisier}, {Mumma}, \& {Weaver}}]{Bockelee-Morvan-cometsII}
{Bockel{\'e}e-Morvan}, D., {Crovisier}, J., {Mumma}, M.~J., \& {Weaver}, H.~A.
  2004, in Comets II, ed. M.~C. {Festou}, H.~U. {Keller}, \& H.~A. {Weaver}
  (University of Arizona Press), 391--423

\bibitem[{{Bodewits} {et~al.}(2011){Bodewits}, {Kelley}, {Li}, {Landsman},
  {Besse}, \& {A'Hearn}}]{Bodewits11}
{Bodewits}, D., {Kelley}, M.~S., {Li}, J.-Y., {et~al.} 2011, \apjl, 733, L3

\bibitem[{{Budzien} {et~al.}(1994){Budzien}, {Festou}, \&
  {Feldman}}]{Budzien94}
{Budzien}, S.~A., {Festou}, M.~C., \& {Feldman}, P.~D. 1994, \icarus, 107, 164

\bibitem[{{Capria} {et~al.}(2012){Capria}, {Marchi}, {de Sanctis}, {Coradini},
  \& {Ammannito}}]{Capria12}
{Capria}, M.~T., {Marchi}, S., {de Sanctis}, M.~C., {Coradini}, A., \&
  {Ammannito}, E. 2012, \aap, 537, A71

\bibitem[{{Cochran} \& {Schleicher}(1993)}]{Cochran+schleicher93}
{Cochran}, A.~L. \& {Schleicher}, D.~G. 1993, \icarus, 105, 235

\bibitem[{{Combi} \& {Delsemme}(1980)}]{Combi+Delsemme80}
{Combi}, M.~R. \& {Delsemme}, A.~H. 1980, \apj, 237, 633

\bibitem[{{de Val-Borro} {et~al.}(2012){de Val-Borro}, {Rezac}, {Hartogh},
  {Biver}, {Bockel{\'e}e-Morvan}, {Crovisier}, {K{\"u}ppers}, {Lis},
  {Szutowicz}, {Blake}, {Emprechtinger}, {Jarchow}, {Jehin}, {Kidger}, {Lara},
  {Lellouch}, {Moreno}, \& {Rengel}}]{devalborro12}
{de Val-Borro}, M., {Rezac}, L., {Hartogh}, P., {et~al.} 2012, \aap, 546, L4

\bibitem[{{Feldman} {et~al.}(2004){Feldman}, {Cochran}, \& {Combi}}]{Feldman04}
{Feldman}, P.~D., {Cochran}, A.~L., \& {Combi}, M.~R. 2004, in Comets II, ed.
  M.~C. {Festou}, H.~U. {Keller}, \& H.~A. {Weaver} (University of Arizona
  Press), 425--447

\bibitem[{{Hadamcik} \& {Levasseur-Regourd}(2009)}]{Hadamcik09}
{Hadamcik}, E. \& {Levasseur-Regourd}, A.~C. 2009, \planss, 57, 1118

\bibitem[{{Hainaut} {et~al.}(2014){Hainaut}, {Boehnhardt}, {Snodgrass},
  {Meech}, {Deller}, {Gillon}, {Jehin}, {Kuehrt}, {Lowry}, {Manfroid},
  {Micheli}, {Mottola}, {Opitom}, {Vincent}, \& {Wainscoat}}]{Hainaut14}
{Hainaut}, O.~R., {Boehnhardt}, H., {Snodgrass}, C., {et~al.} 2014, \aap, 563,
  A75

\bibitem[{Hansen {et~al.}(2017)Hansen, Altwegg, Berthelier, Bieler, Biver,
  Bockel{\'{e}}e-Morvan, Calmonte, Capaccioni, Combi, Keyser, Fiethe, Fougere,
  Fuselier, Gasc, Gombosi, Huang, {Le Roy}, Lee, Nilsson, Rubin, Shou,
  Snodgrass, Tenishev, Toth, Tzou, \& Wedlund}]{Hansen2017}
Hansen, K.~C., Altwegg, K., Berthelier, J.-J., {et~al.} 2017, Monthly Notices
  of the Royal Astronomical Society, in press, stw2413

\bibitem[{{Haser}(1957)}]{Haser57}
{Haser}, L. 1957, Bulletin de la Societe Royale des Sciences de Liege, 43, 740

\bibitem[{{Hsieh} {et~al.}(2014){Hsieh}, {Denneau}, {Fitzsimmons}, {Hainaut},
  {Ishiguro}, {Jedicke}, {Kaluna}, {Keane}, {Kleyna}, {Lacerda}, {MacLennan},
  {Meech}, {Moskovitz}, {Riesen}, {Schunova}, {Snodgrass}, {Trujillo}, {Urban},
  {Vere{\v s}}, {Wainscoat}, \& {Yang}}]{Hsieh14}
{Hsieh}, H.~H., {Denneau}, L., {Fitzsimmons}, A., {et~al.} 2014, \aj, 147, 89

\bibitem[{{Hsieh} {et~al.}(2010){Hsieh}, {Jewitt}, {Lacerda}, {Lowry}, \&
  {Snodgrass}}]{Hsieh10}
{Hsieh}, H.~H., {Jewitt}, D., {Lacerda}, P., {Lowry}, S.~C., \& {Snodgrass}, C.
  2010, \mnras, 403, 363

\bibitem[{{Hsieh} \& {Jewitt}(2006)}]{Hsieh+Jewitt06}
{Hsieh}, H.~H. \& {Jewitt}, D.~C. 2006, Science, 312, 561

\bibitem[{{Hsieh} {et~al.}(2013){Hsieh}, {Kaluna}, {Novakovi{\'c}}, {Yang},
  {Haghighipour}, {Micheli}, {Denneau}, {Fitzsimmons}, {Jedicke}, {Kleyna},
  {Vere{\v s}}, {Wainscoat}, {Ansdell}, {Elliott}, {Keane}, {Meech},
  {Moskovitz}, {Riesen}, {Sheppard}, {Sonnett}, {Tholen}, {Urban}, {Kaiser},
  {Chambers}, {Burgett}, {Magnier}, {Morgan}, \& {Price}}]{Hsieh13}
{Hsieh}, H.~H., {Kaluna}, H.~M., {Novakovi{\'c}}, B., {et~al.} 2013, \apjl,
  771, L1

\bibitem[{{Hsieh} {et~al.}(2011){Hsieh}, {Meech}, \&
  {Pittichov{\'a}}}]{Hsieh11b}
{Hsieh}, H.~H., {Meech}, K.~J., \& {Pittichov{\'a}}, J. 2011, \apjl, 736, L18

\bibitem[{{Hsieh} {et~al.}(2012{\natexlab{a}}){Hsieh}, {Yang}, \&
  {Haghighipour}}]{Hsieh12a}
{Hsieh}, H.~H., {Yang}, B., \& {Haghighipour}, N. 2012{\natexlab{a}}, \apj,
  744, 9

\bibitem[{{Hsieh} {et~al.}(2012{\natexlab{b}}){Hsieh}, {Yang}, {Haghighipour},
  {Kaluna}, {Fitzsimmons}, {Denneau}, {Novakovi{\'c}}, {Jedicke}, {Wainscoat},
  {Armstrong}, {Duddy}, {Lowry}, {Trujillo}, {Micheli}, {Keane}, {Urban},
  {Riesen}, {Meech}, {Abe}, {Cheng}, {Chen}, {Granvik}, {Grav}, {Ip},
  {Kinoshita}, {Kleyna}, {Lacerda}, {Lister}, {Milani}, {Tholen}, {Vere{\v s}},
  {Lisse}, {Kelley}, {Fern{\'a}ndez}, {Bhatt}, {Sahu}, {Kaiser}, {Chambers},
  {Hodapp}, {Magnier}, {Price}, \& {Tonry}}]{Hsieh12b}
{Hsieh}, H.~H., {Yang}, B., {Haghighipour}, N., {et~al.} 2012{\natexlab{b}},
  \apjl, 748, L15

\bibitem[{{Hsieh} {et~al.}(2012{\natexlab{c}}){Hsieh}, {Yang}, {Haghighipour},
  {Novakovi{\'c}}, {Jedicke}, {Wainscoat}, {Denneau}, {Abe}, {Chen},
  {Fitzsimmons}, {Granvik}, {Grav}, {Ip}, {Kaluna}, {Kinoshita}, {Kleyna},
  {Knight}, {Lacerda}, {Lisse}, {Maclennan}, {Meech}, {Micheli}, {Milani},
  {Pittichov{\'a}}, {Schunova}, {Tholen}, {Wasserman}, {Burgett}, {Chambers},
  {Heasley}, {Kaiser}, {Magnier}, {Morgan}, {Price}, {J{\o}rgensen}, {Dominik},
  {Hinse}, {Sahu}, \& {Snodgrass}}]{Hsieh12c}
{Hsieh}, H.~H., {Yang}, B., {Haghighipour}, N., {et~al.} 2012{\natexlab{c}},
  \aj, 143, 104

\bibitem[{{Huebner} {et~al.}(1992){Huebner}, {Keady}, \& {Lyon}}]{Huebner92}
{Huebner}, W.~F., {Keady}, J.~J., \& {Lyon}, S.~P. 1992, \apss, 195, 1

\bibitem[{{Ishiguro} {et~al.}(2011){Ishiguro}, {Hanayama}, {Hasegawa},
  {Sarugaku}, {Watanabe}, {Fujiwara}, {Terada}, {Hsieh}, {Vaubaillon}, {Kawai},
  {Yanagisawa}, {Kuroda}, {Miyaji}, {Fukushima}, {Ohta}, {Hamanowa}, {Kim},
  {Pyo}, \& {Nakamura}}]{Ishiguro11a}
{Ishiguro}, M., {Hanayama}, H., {Hasegawa}, S., {et~al.} 2011, \apjl, 740, L11

\bibitem[{{Jewitt}(2012)}]{Jewitt12review}
{Jewitt}, D. 2012, \aj, 143, 66

\bibitem[{{Jewitt} {et~al.}(2014){Jewitt}, {Agarwal}, {Li}, {Weaver},
  {Mutchler}, \& {Larson}}]{Jewitt14a}
{Jewitt}, D., {Agarwal}, J., {Li}, J., {et~al.} 2014, \apjl, 784, L8

\bibitem[{Jewitt {et~al.}(2015)Jewitt, Agarwal, Peixinho, Weaver, Mutchler,
  Hui, Li, \& Larson}]{Jewitt2015g}
Jewitt, D., Agarwal, J., Peixinho, N., {et~al.} 2015, The Astronomical Journal,
  149, 81

\bibitem[{{Jewitt} {et~al.}(2013){Jewitt}, {Agarwal}, {Weaver}, {Mutchler}, \&
  {Larson}}]{Jewitt13d}
{Jewitt}, D., {Agarwal}, J., {Weaver}, H., {Mutchler}, M., \& {Larson}, S.
  2013, \apjl, 778, L21

\bibitem[{{Jewitt} {et~al.}(2009){Jewitt}, {Yang}, \&
  {Haghighipour}}]{Jewitt09}
{Jewitt}, D., {Yang}, B., \& {Haghighipour}, N. 2009, \aj, 137, 4313

\bibitem[{Kelley {et~al.}(2016)Kelley, Woodward, Bodewits, Farnham, Gudipati,
  Harker, Hines, Knight, Kolokolova, Li, de~Pater, Protopapa, Russell, Sitko,
  \& Wooden}]{kelley16}
Kelley, M. S.~P., Woodward, C.~E., Bodewits, D., {et~al.} 2016, Publications of
  the Astronomical Society of the Pacific, 128, 018009

\bibitem[{{Kolokolova} {et~al.}(2004){Kolokolova}, {Hanner},
  {Levasseur-Regourd}, \& {Gustafson}}]{Kolokolova2004}
{Kolokolova}, L., {Hanner}, M.~S., {Levasseur-Regourd}, A.-C., \& {Gustafson},
  B.~{\AA}.~S. 2004, {Physical properties of cometary dust from light
  scattering and thermal emission}, ed. M.~C. {Festou}, H.~U. {Keller}, \&
  H.~A. {Weaver}, 577--604

\bibitem[{{Lee} {et~al.}(2015){Lee}, {von Allmen}, {Allen}, {Beaudin}, {Biver},
  {Bockel{\'e}e-Morvan}, {Choukroun}, {Crovisier}, {Encrenaz}, {Frerking},
  {Gulkis}, {Hartogh}, {Hofstadter}, {Ip}, {Janssen}, {Jarchow}, {Keihm},
  {Lellouch}, {Leyrat}, {Rezac}, {Schloerb}, {Spilker}, {Gaskell}, {Jorda},
  {Keller}, \& {Sierks}}]{Lee15}
{Lee}, S., {von Allmen}, P., {Allen}, M., {et~al.} 2015, \aap, 583, A5

\bibitem[{{Licandro} {et~al.}(2011){Licandro}, {Campins}, {Tozzi}, {de
  Le{\'o}n}, {Pinilla-Alonso}, {Boehnhardt}, \& {Hainaut}}]{Licandro11b}
{Licandro}, J., {Campins}, H., {Tozzi}, G.~P., {et~al.} 2011, \aap, 532, A65

\bibitem[{{Licandro} {et~al.}(2013){Licandro}, {Moreno}, {de Le{\'o}n},
  {Tozzi}, {Lara}, \& {Cabrera-Lavers}}]{Licandro13}
{Licandro}, J., {Moreno}, F., {de Le{\'o}n}, J., {et~al.} 2013, \aap, 550, A17

\bibitem[{{MacLennan} \& {Hsieh}(2012)}]{MacLennan+Hsieh2012}
{MacLennan}, E.~M. \& {Hsieh}, H.~H. 2012, \apjl, 758, L3

\bibitem[{{Mainzer} {et~al.}(2011){Mainzer}, {Grav}, {Masiero}, {Hand},
  {Bauer}, {Tholen}, {McMillan}, {Spahr}, {Cutri}, {Wright}, {Watkins}, {Mo},
  \& {Maleszewski}}]{Mainzer11}
{Mainzer}, A., {Grav}, T., {Masiero}, J., {et~al.} 2011, \apj, 741, 90

\bibitem[{{Meech} {et~al.}(2011){Meech}, {Pittichov{\'a}}, {Yang}, {Zenn},
  {Belton}, {A'Hearn}, {Bagnulo}, {Bai}, {Barrera}, {Bauer}, {Bedient},
  {Bhatt}, {Boehnhardt}, {Brosch}, {Buie}, {Candia}, {Chen}, {Chesley},
  {Chiang}, {Choi}, {Cochran}, {Duddy}, {Farnham}, {Fern{\'a}ndez},
  {Guti{\'e}rrez}, {Hainaut}, {Hampton}, {Herrmann}, {Hsieh}, {Kadooka},
  {Kaluna}, {Keane}, {Kim}, {Kleyna}, {Krisciunas}, {Lauer}, {Lara},
  {Licandro}, {Lowry}, {McFadden}, {Moskovitz}, {Mueller}, {Polishook}, {Raja},
  {Riesen}, {Sahu}, {Samarasinha}, {Sarid}, {Sekiguchi}, {Sonnett}, {Suntzeff},
  {Taylor}, {Tozzi}, {Vasundhara}, {Vincent}, {Wasserman}, {Webster-Schultz},
  \& {Zhao}}]{Meech11}
{Meech}, K.~J., {Pittichov{\'a}}, J., {Yang}, B., {et~al.} 2011, \icarus, 213,
  323

\bibitem[{Modigliani {et~al.}(2010)Modigliani, Goldoni, Royer, Haigron,
  Guglielmi, Fran{\c{c}}ois, Horrobin, Bristow, Vernet, Moehler, Kerber,
  Ballester, Mason, \& Christensen}]{XSHOOTER-pipeline}
Modigliani, A., Goldoni, P., Royer, F., {et~al.} 2010, in Society of
  Photo-Optical Instrumentation Engineers (SPIE) Conference Series, Vol. 7737,
  Society of Photo-Optical Instrumentation Engineers (SPIE) Conference Series,
  ed. D.~R. Silva, A.~B. Peck, \& B.~T. Soifer, 773728

\bibitem[{{Moreno} {et~al.}(2013){Moreno}, {Cabrera-Lavers}, {Vaduvescu},
  {Licandro}, \& {Pozuelos}}]{Moreno13}
{Moreno}, F., {Cabrera-Lavers}, A., {Vaduvescu}, O., {Licandro}, J., \&
  {Pozuelos}, F. 2013, \apjl, 770, L30

\bibitem[{{O'Rourke} {et~al.}(2013){O'Rourke}, {Snodgrass}, {de Val-Borro},
  {Biver}, {Bockel{\'e}e-Morvan}, {Hsieh}, {Teyssier}, {Fernandez}, {Kueppers},
  {Micheli}, \& {Hartogh}}]{ORourke13}
{O'Rourke}, L., {Snodgrass}, C., {de Val-Borro}, M., {et~al.} 2013, \apjl, 774,
  L13

\bibitem[{{Prialnik} {et~al.}(2004){Prialnik}, {Benkhoff}, \&
  {Podolak}}]{Prialnik04}
{Prialnik}, D., {Benkhoff}, J., \& {Podolak}, M. 2004, in Comets II, ed. M.~C.
  {Festou}, H.~U. {Keller}, \& H.~A. {Weaver} (University of Arizona Press),
  359--387

\bibitem[{{Prialnik} \& {Rosenberg}(2009)}]{Prialnik09}
{Prialnik}, D. \& {Rosenberg}, E.~D. 2009, \mnras, 399, L79

\bibitem[{{Rotundi} {et~al.}(2015){Rotundi}, {Sierks}, {Della Corte}, {Fulle},
  {Gutierrez}, {Lara}, {Barbieri}, {Lamy}, {Rodrigo}, {Koschny}, {Rickman},
  {Keller}, {L{\'o}pez-Moreno}, {Accolla}, {Agarwal}, {A'Hearn}, {Altobelli},
  {Angrilli}, {Barucci}, {Bertaux}, {Bertini}, {Bodewits}, {Bussoletti},
  {Colangeli}, {Cosi}, {Cremonese}, {Crifo}, {Da Deppo}, {Davidsson}, {Debei},
  {De Cecco}, {Esposito}, {Ferrari}, {Fornasier}, {Giovane}, {Gustafson},
  {Green}, {Groussin}, {Gr{\"u}n}, {G{\"u}ttler}, {Herranz}, {Hviid}, {Ip},
  {Ivanovski}, {Jer{\'o}nimo}, {Jorda}, {Knollenberg}, {Kramm}, {K{\"u}hrt},
  {K{\"u}ppers}, {Lazzarin}, {Leese}, {L{\'o}pez-Jim{\'e}nez}, {Lucarelli},
  {Lowry}, {Marzari}, {Epifani}, {McDonnell}, {Mennella}, {Michalik}, {Molina},
  {Morales}, {Moreno}, {Mottola}, {Naletto}, {Oklay}, {Ortiz}, {Palomba},
  {Palumbo}, {Perrin}, {Rodr{\'{\i}}guez}, {Sabau}, {Snodgrass}, {Sordini},
  {Thomas}, {Tubiana}, {Vincent}, {Weissman}, {Wenzel}, {Zakharov}, \&
  {Zarnecki}}]{Rotundi2015Sci}
{Rotundi}, A., {Sierks}, H., {Della Corte}, V., {et~al.} 2015, Science, 347,
  3905

\bibitem[{{Schleicher}(2010)}]{Schleicher10}
{Schleicher}, D.~G. 2010, \aj, 140, 973

\bibitem[{{Schleicher} \& {A'Hearn}(1988)}]{Schleicher+AHearn88}
{Schleicher}, D.~G. \& {A'Hearn}, M.~F. 1988, \apj, 331, 1058

\bibitem[{Snodgrass {et~al.}(2016)Snodgrass, Jehin, Manfroid, Opitom,
  Fitzsimmons, Tozzi, Faggi, Yang, Knight, Conn, Lister, Hainaut, Bramich,
  Lowry, Rozek, Tubiana, \& Guilbert-Lepoutre}]{Snodgrass2016}
Snodgrass, C., Jehin, E., Manfroid, J., {et~al.} 2016, Astronomy {\&}
  Astrophysics, 588, A80

\bibitem[{{Snodgrass} {et~al.}(2010){Snodgrass}, {Tubiana}, {Vincent},
  {Sierks}, {Hviid}, {Moissi}, {Boehnhardt}, {Barbieri}, {Koschny}, {Lamy},
  {Rickman}, {Rodrigo}, {Carry}, {Lowry}, {Laird}, {Weissman}, {Fitzsimmons},
  {Marchi}, \& {OSIRIS Team}}]{Snodgrass10b}
{Snodgrass}, C., {Tubiana}, C., {Vincent}, J.-B., {et~al.} 2010, \nat, 467, 814

\bibitem[{{Stevenson} {et~al.}(2012){Stevenson}, {Kramer}, {Bauer}, {Masiero},
  \& {Mainzer}}]{Stevenson12}
{Stevenson}, R., {Kramer}, E.~A., {Bauer}, J.~M., {Masiero}, J.~R., \&
  {Mainzer}, A.~K. 2012, \apj, 759, 142

\bibitem[{{Tholen} {et~al.}(2012){Tholen}, {Elliott}, {Sato}, {Buzzi},
  {Devore}, {Foglia}, {Vorobjov}, {Holmes}, \& {Williams}}]{Tholen12}
{Tholen}, D.~J., {Elliott}, T., {Sato}, H., {et~al.} 2012, Central Bureau
  Electronic Telegrams, 3252, 1

\bibitem[{{Tseng} {et~al.}(2007){Tseng}, {Bockel{\'e}e-Morvan}, {Crovisier},
  {Colom}, \& {Ip}}]{Tseng07}
{Tseng}, W.-L., {Bockel{\'e}e-Morvan}, D., {Crovisier}, J., {Colom}, P., \&
  {Ip}, W.-H. 2007, \aap, 467, 729

\bibitem[{{Vernet} {et~al.}(2011){Vernet}, {Dekker}, {D'Odorico}, {Kaper},
  {Kjaergaard}, {Hammer}, {Randich}, {Zerbi}, {Groot}, {Hjorth}, {Guinouard},
  {Navarro}, {Adolfse}, {Albers}, {Amans}, {Andersen}, {Andersen}, {Binetruy},
  {Bristow}, {Castillo}, {Chemla}, {Christensen}, {Conconi}, {Conzelmann},
  {Dam}, {de Caprio}, {de Ugarte Postigo}, {Delabre}, {di Marcantonio},
  {Downing}, {Elswijk}, {Finger}, {Fischer}, {Flores}, {Fran{\c c}ois},
  {Goldoni}, {Guglielmi}, {Haigron}, {Hanenburg}, {Hendriks}, {Horrobin},
  {Horville}, {Jessen}, {Kerber}, {Kern}, {Kiekebusch}, {Kleszcz}, {Klougart},
  {Kragt}, {Larsen}, {Lizon}, {Lucuix}, {Mainieri}, {Manuputy}, {Martayan},
  {Mason}, {Mazzoleni}, {Michaelsen}, {Modigliani}, {Moehler}, {M{\o}ller},
  {Norup S{\o}rensen}, {N{\o}rregaard}, {P{\'e}roux}, {Patat}, {Pena}, {Pragt},
  {Reinero}, {Rigal}, {Riva}, {Roelfsema}, {Royer}, {Sacco}, {Santin},
  {Schoenmaker}, {Spano}, {Sweers}, {Ter Horst}, {Tintori}, {Tromp}, {van
  Dael}, {van der Vliet}, {Venema}, {Vidali}, {Vinther}, {Vola}, {Winters},
  {Wistisen}, {Wulterkens}, \& {Zacchei}}]{X-SHOOTER}
{Vernet}, J., {Dekker}, H., {D'Odorico}, S., {et~al.} 2011, \aap, 536, A105

\bibitem[{{Yang} \& {Hsieh}(2011)}]{Yang11}
{Yang}, B. \& {Hsieh}, H. 2011, \apjl, 737, L39

\end{thebibliography}
\end{document}